\documentstyle[aps,twocolumn]{revtex}
\def\opone{\leavevmode\hbox{\small1\kern-3.8pt\normalsize1}}
\def\sign{{\rm sign}}
\newcommand{\beq}{\begin{equation}}
\newcommand{\eeq}{\end{equation}}
\newcommand{\beqa}{\begin{eqnarray}}
\newcommand{\eeqa}{\end{eqnarray}}
\newcommand{\ban}{\begin{eqnarray*}}
\newcommand{\ean}{\end{eqnarray*}}

\newcommand{\one}{\leavevmode\hbox{\small1\normalsize\kern-.33em1}}

\begin{document}

\title{A local variable model for entanglement swapping exploiting the detection loophole}

\author{Nicolas Gisin and Bernard Gisin\\
Group of Applied Physics, University of Geneva\\
20, rue de l'Ecole-de-M\'edecine, CH-1211 Geneva 4, Switzerland}
%\date{}
\maketitle

\begin{abstract}
In an entanglement swapping process two initially uncorrelated qubits become entangled, without any
direct interaction. We present a model using local variables aiming at reproducing this
remarkable process, under the realistic assumption of finite detection efficiencies. 
The model assumes that the local variables describing the two qubits are
initially completely uncorrelated. Nevertheless, we show that once conditioned on the Bell
measurement result, the local variables bear enough correlation to simulate quantum measurement
results with correlation very close to the quantum prediction. When only a partial Bell
measurement is simulated, as carried out is all experiments so far, then the model recovers
analytically the quantum prediction.
\end{abstract}

%\vspace{1cm}

%Keywords:

\vspace{1cm}

\section{Introduction}
%=====================
Quantum theory is nonlocal and the richness of entanglement impressive.
For the simplest possible case of a two qubits system, the states can be
divided into:
\begin{enumerate} 
\item the separable ones: states that can be prepared locally and whose correlation can be
described locally, 
\item the poorly entangled ones: states that can't be prepared locally but which admit a local
hidden variable model, i.e. the correlation between any projective measurement can be
described locally,
\item the fully entangled ones: states that can't be prepared locally and whose
correlation - even between projective measurements - can't be described locally.
\end{enumerate}
To this list one should add the class of states useful for teleportation \cite{BarrettTelep} and the ones that can
be distilled (using 1- or 2-way classical communication). One might also need to consider 
generalized measurements (POVMs) \cite{BarrettPOVM}. 
In higher dimensions it is known that these structures are even more involved.
Similarly, when considering systems with more than two subsystems further classes are
necessary to classify the states (i.e GHZ- and W-states for 3 qubits \cite{Wstate}).

In this letter we consider the class of states that can be experimentally demonstrated to be
nonlocal. This class depends on the efficiency of available detectors. The problem may thus
seem independent from the entanglement classification program, but this is not really so
for at least two reasons. First, this program being part of physics, it must lead to experimental tests.
For example, the possibility to experimentally demonstrate entanglement is probably the best --
if not the only -- way to convince a doubtful customer that a pair of devices constitute the 
source and receiver of a truly quantum cryptographic system (i.e. that they are really quantum and
not merely fakes!) \cite{MayersLao,GisinRMP}. Quantum error correction and linear optics quantum
computation provide further examples: for each such
schemes to operate, each component must surpass a certain threshold in efficiency. A second
motivation for the question we address is its connection to the timely problem of communication
complexity \cite{BrassardComComplexity}. Indeed, every local model mimicking quantum
correlation for a detection efficiency $\eta$ can be associated to a model for perfect
detectors but for a more noisy state: a mixture of the original
state and of a fraction of the maximally mixed state. 
The association works as follows: in each instance where the original model assumes that
the detector fails to register a count, the corresponding model assumes that a random
(i.e. uncorrelated) output is produced\footnote{This applies also to the model we
presented in a previous publication\cite{Gisinlhv}. In this case, either one detector has an efficiency
limited to 50\%, or the maximally entangled 2-qubit state is mixed with 50\% of the maximally
entangled state, i.e. is the Werner state. At the time of writing our previous model, we
where not aware of the close connection between our model and Werner's local hidden variable
model.}.

For 2-qubit systems, the problem of demonstrating quantum non locally via violation of the
Bell inequality has a long history. Pearle, Shimony and Clauser seem to have been the
first to notice the {\it detection loophole}, latter studied in details by Santos and others
\cite{detLoophole}. This loophole resisted to experimenters until
last year when the Boulder group published results obtains with detectors
sufficiently efficient to close it (but with only 3 microns separation, so that quantum
nonlocality is still not demonstrated as beautifully as one would like for such a 
fundamental concept!) \cite{Wineland}.

This letter is organized as follows. First, in the next section, we briefly review the
detection loophole and how a simple local hidden variable model can exploit it. Next, we
present the basic intuition behind this letter: in entanglement swapping \cite{entswap} it is very
natural to assume that the two input photon pairs are independent (recall that, in
principle they are produced by two independent sources). Consequently, any assumed local
variables associated to the qubits, after the entangled swapping process has completed, must be
uncorrelated, or at least only poorly correlated. This reduced correlation should make it 
more difficult to simulate the quantum correlation, hence should make it easier
to close the detection loophole. This intuition is expanded in section \ref{intuition}.
Finally, two explicit models are presented and analysed. The first model assumes
a partial Bell measurement (section \ref{firstModel}), the second one a complete
Bell measurement - but always with a finite detection efficiency (section \ref{secondModel}).

\section{Reminder of the detection loophole and a related simple model of quantum correlation}\label{lhvmodel}
%=============================================================================================
The idea behind the detection loophole is very simple and natural. It merely states that the
probability that a particle is detected depends, among others, on the particle's state. This is
true as well in classical as in quantum physics. There is thus no reason to reject the 
assumption that the same would be true in some hypothetical theory which would incorporate additional
variables $\lambda$'s, whether local or not. But then, if the additional variables $\lambda$
are not under the physicist's control (i.e. are hidden), the detection probability may vary from one run of the
experiment to the next. This simple idea allows one to construct a straightforward and explicit model
reproducing the quantum correlation exactly, using only local variables\cite{Gisinlhv}.

It is useful to remind this simple model here, since the entanglement swapping models of the next
sections are strongly inspired by it (but for the proofs we refer the reader to \cite{Gisinlhv}).
The model uses the Poincar\'e sphere $S^2$ representation of the qubit state space. Here we remind
the simplest model, reproducing the correlation of the quantum singlet state. This model treats
the two qubits asymmetrically, but it is straightforward to symmetries the model\cite{Gisinlhv}. 
The jth qubit's state is described by a normalized vector $\vec\lambda_j$, j=1,2. When the first
qubit is measured along direction $\vec a$ (i.e. the operator representing the measured quantity
is $\vec a\vec\sigma$, with $\vec\sigma$ the Pauli matrices), 
then the probability that the particle carrying the qubit is not
detected is $1-|\vec a\vec\lambda_1|$, and if the particle is detected, then the outcome is 
\sign($\vec a\vec\lambda_1$). When the second qubit is measured along direction $\vec b$, it
always produces an outcome equal to \sign($\vec b\vec\lambda_2$). Assuming that the qubit pair
source produces random $\vec\lambda_1$, but with $\vec\lambda_2=-\vec\lambda_1$, this model predicts a mean
detection efficiency (averaged over all $\vec\lambda$'s and over all detectors) of 0.75 and a
correlation (conditioned on both qubits producing an outcome) equal to the quantum
prediction: $E(\vec a,\vec b)=-\vec a\vec b$ ! This rather simple and natural (in the spirit
of local hidden variables) model proves that no experiments on two maximally entangled 
qubits with detector efficiencies
smaller than 75\% can definitively demonstrate quantum nonlocality, whatever the number of
settings and measurements used on each side (the CH inequality \cite{CH} proves that when restricted to two
settings on each side, the minimal detection efficiency using qubits in the singlet state is
even slightly larger\footnote{Whether this difference means that there is a better
local model or a better Bell inequality - e.g. involving more than two settings by site or
using generalized measurements - is an interesting open question.}: 82.8\%).

\section{The intuition}\label{intuition}
%======================
The task of an experimental demonstration of quantum nonlocality without loopholes requires
to surpass some technological thresholds. There is little doubt that this will be achieved in the
future, but there is clearly also the desire to simplify the experiment by clever reasoning
\cite{actpassDetloophole}. One possibility is to go to higher dimensional systems. Indeed, Serge Massar
showed that the required detection efficiency decreases exponentially with the dimension
when two system are maximally entangled\cite{MassarDetLoophole}. Here we explore another idea,
related to entanglement swapping\cite{entswap}. This very remarkable process allows two qubits that
never interacted to become entangled via an interaction which affects only two other qubits, each
of the latter being initially entangled with one of the former qubits, see figure 1. When
trying to simulate this remarkable process with local physics, it is very natural to assume that
the local variables $\lambda_1=-\lambda_2$ characterizing the first entangled qubit pair is independent
of the local variable $\lambda_3=-\lambda_4$ characterizing the second entangled qubit pair. Consequently,
after the entanglement swapping, i.e. after one qubit of each pair (e.g. qubits number 2 and 3) 
interacted in a measurement
process, the two remaining qubits are characterized by $\lambda_1$ and $\lambda_4$,
respectively. Since $\lambda_1$ and $\lambda_4$ are independent, the two remaining qubits are
uncorrelated, in agreement with the quantum prediction. However, quantum mechanics predicts
that when the state of qubits number 1 and 4 is conditioned on the result of the measurement carried out on their
twins (i.e. on qubits 2 and 3), then the qubits 1 and 4 are fully entangled. Since the measurement has 4 possibly outcomes,
the set of variables  $(\lambda_2,\lambda_3)$ must be divided into 5 subsets: 4 subsets
corresponding to the 4 possible measurement outcomes and the fifth one to the case of no-detection.
In each subset $\lambda_2$ and $\lambda_3$ may have some correlation, but clearly not full
correlation, in particular not of the form $\lambda_2=-\lambda_1$ as in the simple
model recalled in the previous section. Hence, it is intuitively clear that an experiment can rule out such local models
with lower detection efficiencies than required in standard Bell tests. But we shall see that
the improvement is surprisingly small.

\section{A first model: case of partial Bell-measurement}\label{firstModel}
%==============================================================
In this first model of entanglement swapping we assume that only an incomplete Bell
measurement is performed on the two central qubits number 2 and 3, see figure 1. This
partial Bell measurement only singles out the singled state, as is the case in 
all experiments today \cite{entswapexp}. In this model each of the 4 qubit states are described by
normalized vectors $\vec\lambda_j$, j=1,2,3,4. The two qubit pair sources produce random
$\vec\lambda$'s, but with $\vec\lambda_2=-\vec\lambda_1$ and $\vec\lambda_4=-\vec\lambda_3$, in
agreement with the simple model recalled in section \ref{lhvmodel}. The first entanglement
swapping model assumes that the partial Bell measurement produces the singlet-outcome iff
\beq
\vec\lambda_2\cdot\vec\lambda_3\le1-\frac{\eta^2}{2},
\eeq
where $\eta$ is a parameter of the model
characterizing the averaged detection efficiency ($0\le\eta\le1$). Accordingly, the 
probability of the singlet-outcome reads:
\beqa
P(singlet)&=&\int_{S^2}\int_{S^2}\frac{d\vec\lambda_2d\vec\lambda_3}{(4\pi)^2}~\theta(1-\frac{\eta^2}{2}-\vec\lambda_2\vec\lambda_3) \\
&=&\int_{-1}^1\frac{dz}{2}~\theta(1-\frac{\eta^2}{2}-z) \\
&=&\frac{\eta^2}{4}
\eeqa
where $\theta$ is Heavyside's function. Hence, as announced, the model predicts that the
partial Bell measurement produces the outcome {\it singlet} with the same probability as
quantum mechanics when both qubits are detected with an efficiency $\eta$ (the factor $\frac{1}{4}$
takes into account that in a partial Bell measurement only 1 of the 4 Bell states is detected,
even in principle). Now, the model is exactly as in section \ref{lhvmodel}:
when the first qubit is measured along direction $\vec a$ it produces the result $\sign(\vec a
\vec\lambda_1)$ with probability $|\vec a\vec\lambda_1|$ and no outcome at all with the
complementary probability. The last qubit, when measured along $\vec b$ always produces the
result $\sign(\vec b\vec\lambda_4)$. Let us stress that the asymmetry between qubit 1 and 4
in this model can be remove by adding auxiliary random variables, similarly to \cite{Gisinlhv}.

Let us now compute the predictions of this model. First, the probability that the
partial Bell measurement produces the outcome {\it singlet} and that the first qubit
produces an outcome, i.e the probability that all the 4 qubits fire a detector, reads
(using $\vec\lambda_2=-\vec\lambda_1$):
\beqa
p&\equiv& Prob(Bell=singlet~\&~a\ne0) \nonumber\\
&=&\int_{S^2}\int_{S^2}\frac{d\vec\lambda_1d\vec\lambda_3}{(4\pi)^2}|\vec a\vec\lambda_1|\theta\left(\vec\lambda_1\vec\lambda_3-(1-\frac{\eta^2}{2})\right) \\
&=&\int_{S^2}\frac{d\vec\lambda_1}{4\pi}|\vec a\vec\lambda_1|\frac{1-(1-\eta^2/2)}{2} \\
&=&\frac{\eta^2}{8}
\eeqa
Next, the distribution probability of $\vec\lambda_1$ and $\vec\lambda_4$ conditioned of all
qubits being detected reads:
\beqa
\rho(\vec\lambda_1,\vec\lambda_4|Bell=singlet~\&~a\ne0)= \\
=\frac{\frac{1}{(4\pi)^2}|\vec a\vec\lambda_1|\theta(1-\frac{\eta^2}{2}-\vec\lambda_1\vec\lambda_4)}{p}
\eeqa
Finally, the correlation function when qubits 1 and 4 are measured along directions 
$\vec a$ and $\vec b$, conditioned on all qubits being detected, reads:
\beqa
E(\vec a,\vec b)&=&\int_{S^2}\int_{S^2}d\vec\lambda_1d\vec\lambda_4 
~\rho(\vec\lambda_1,\vec\lambda_4|Bell=singlet~\&~a\ne0)\cdot \nonumber\\
&&\hspace{15 mm}\cdot\sign(\vec a\vec\lambda_1)\sign(\vec b\vec\lambda_4) \\
&=&\frac{1}{2\pi^2\eta^2}\int_{S^2}\int_{S^2}d\vec\lambda_1d\vec\lambda_4
~\theta(1-\frac{\eta^2}{2}-\vec\lambda_1\vec\lambda_4)\cdot \nonumber\\
&&\hspace{15 mm}\cdot\vec a\vec\lambda_1\sign(\vec b\vec\lambda_4) \\
&=&\frac{1}{2\pi^2\eta^2}\int_{S^2}d\vec\lambda_4~\sign(\vec b\vec\lambda_4)\cdot \nonumber\\
&&\hspace{10mm}\cdot\int_{S^2}d\vec\lambda_1\theta(1-\frac{\eta^2}{2}-\vec\lambda_1\vec\lambda_4)
\vec a\vec\lambda_1 \\
&=&\frac{1-(1-\eta^2/2)^2}{2\pi\eta^2}\int_{S^2}d\vec\lambda_4~\sign(\vec b\vec\lambda_4)\vec a\vec\lambda_4 \\
&=& -(1-\frac{\eta^2}{4})\vec a\vec b
\eeqa

This model of entanglement swapping with a partial Bell measurement predicts thus 
correlations of exactly the same form as the quantum prediction, but with a reduced
visibility
\beq
V=1-\frac{\eta^2}{4}
\label{Vmin}
\eeq
Consequently - and surprisingly! - using entanglement
swapping with a partial Bell measurement barely helps demonstrating quantum nonlocality.
Indeed, one would need to obtain a measured correlation between the initially independent
two qubits with a visibility larger than 75\%, i.e. larger than the usual Bell-threshold
of 70.7\% (this usual threshold does not take into account the detector loophole which we
consider here). Actually, with realistic Bell measurements of finite efficiency, the
required visibility is even higher, as indicated in (\ref{Vmin})! For example, for
realistic photon counters with $\eta\approx0.4$, one has $V\approx 94\%$.

Let us stress the surprising feature of this model. Initially the two local variables
$\vec\lambda_1$ and $\vec\lambda_4$ are completely independent. Even when conditioned upon the
partial Bell measurement outcome {\it singlet}, these two local variables are only poorly
correlated. Nevertheless the correlation predicted from these poorly correlated
local variables has the quantum mechanical shape, with merely a reduced visibility (note that
this model, like the one in section \ref{lhvmodel}, assumes a mean detection efficiency smaller
or equal to 75\%).
Moreover, for realistic detection efficiencies the visibility is only slightly
below 1. Consequently, no experiments on entanglement swapping experiments with partial Bell 
measurement and realistic detector efficiencies can definitively demonstrate quantum
nonlocality, whatever the number of settings and measurements  used on each side!

\section{Second model: case of complete Bell-measurements}\label{secondModel}
%=========================================================
In the light of the rather disappointing result of the previous section, it is tempting to
investigate the case of entanglement swapping with a complete Bell measurement. Indeed,
it is intuitively clear that in such a case the Bell measurement can't correlate the
local variables $\vec\lambda_1$ and $\vec\lambda_4$ as much as in the first model.
The idea of this second model is closely inspired by the previous one and by the well known
fact that all the 4 Bell states can be obtained from the singlet one, $\psi^{(-)}$, with only one local
rotations:
\beqa
\phi^{(-)}&=&\opone\otimes\sigma_x~\psi^{(-)} \\
\phi^{(+)}&=&\opone\otimes\sigma_y~\psi^{(-)} \\
\psi^{(+)}&=&\opone\otimes\sigma_z~\psi^{(-)} \\
\eeqa
Hence, the idea is that the Bell measurement results should produce the following correlation:
\beqa
\psi^{(-)}&\Rightarrow& \vec\lambda_1\approx-\vec\lambda_4 \\
\phi^{(-)}&\Rightarrow& \vec\lambda_1\approx- R_x\vec\lambda_4 \\
\phi^{(+)}&\Rightarrow& \vec\lambda_1\approx- R_y\vec\lambda_4 \\
\psi^{(+)}&\Rightarrow& \vec\lambda_1\approx- R_z\vec\lambda_4 
\eeqa
where $R_k$ represents a $\pi$-rotation around the k-axis. Following this idea,
the second model assumes that the Bell measurement result is the one corresponding to the
most negative scalar product among the following four: $\vec\lambda_1\cdot\vec\lambda_4,~
\vec\lambda_1\cdot R_x\vec\lambda_4,~\vec\lambda_1\cdot R_y\vec\lambda_4$ and 
$\vec\lambda_1\cdot R_z\vec\lambda_4$, provided it is lower than -{\it Limit}, where
{\it Limit} is a parameter which determines the Bell measurement efficiency, see Fig. 2.

For this second model, we were unable to find the analytic form of the predicted correlation.
However, simple numerical simulations can be used. Fig. 3 presents the mean fidelity
as estimated by the correlation when both analysers measuring qubits 1 and 4 are parallel.
Finally, Fig. 4 and 5 illustrate the simulated correlation when Alice's analyser is fixed and
Bob's one rotated. On each graph there are 4 correlation corresponding to the 2x2 possible
results on Alice and Bob sites together with 4 sinusoidal fits.
These numerical results, and many other not shown here, demonstrate that
the shape of the correlation is no longer strictly proportional to the quantum one:
$E(\vec a,\vec b)\ne cst\cdot\vec a\vec b$. However, the difference is so small that it is
still essential impossible to find a measurable difference!

\section{Conclusion}
%==================
Quantum nonlocality is fundamental for the wordview offered by physics and central for
several applications of quantum information theory. Consequently, it deserves to be carefully
tested. The detection loophole is annoying because it resists almost all experimental efforts
(see however \cite{Wineland}). It is also fascinating because of the connections mentioned
in the introduction with quantum cryptography and with quantum communication complexity: it
offer a nice example of the close connection between basic and applied physics. In this letter
we have shown that even for a process like entanglement swapping, where entanglement is
established between particle that never interacted directly, the detection loophole remains
exceedingly difficult to close.

\section*{Acknowledgments}
%========================
Support by the Swiss FNRS and NCCR {\it Quantum Photonics} are acknowledged.

\section*{Figure Captions}
\begin{enumerate}
\item General scheme of the entanglement swapping setup. Each 2-photon source produces qubit pairs
in the singlet state. In the local variable model under consideration the state of each qubit is 
entirely characterized by a normalized 3-dim vector, i.e. a point on the Poincar\'e sphere,
each randomly distributed on the entire sphere. The
sources produce correlated qubits with local parameters $\vec\lambda_2=-\vec\lambda_1$ and
$\vec\lambda_4=-\vec\lambda_3$. The 2 sources are independent, consequently $\vec\lambda_1$ and
$\vec\lambda_4$ are statistically independent. The Bell measurement produces one of 5 possible
outcomes: one of the 4 usual Bell states or the {\it no-detection} result. In the local variable model
the probability for these 5 outcomes depend only on $\vec\lambda_2$ and $\vec\lambda_3$. In the
simplified model (section \ref{firstModel}) only a partial Bell measurement is considered, i.e. only 2
outcomes are considered: either {\it singlet}, or  {\it no-detection}.

\item Numerical simulation of the second model, assuming a complete Bell measurement
(on $10^8$ samples). The probability that the Bell measurement gives a result is computed in 
function of the parameter {\it Limit} (see text).

\item The fidelity of the entanglement swapping simulation is computed (on $10^7$ samples)
in function of the probability that the complete Bell measurement gives a result.

\item The 4 correlation function between the results on Alice and on Bob sites (two possible outcomes
per site) are computed (on $10^6$ samples)
in function of a rotation of Bob's analyser. The efficiency of the Bell measurement is
assumed at $\approx 50\%$. Four sinusoidal
fits are also displayed, illustrating that the model, although different from the quantum
prediction, is in practice undistinguishable.

\item Same as 4, but for different analyser setting and a Bell measurement efficiency 
$\approx 90\%$.

\end{enumerate}

\end{document}